# WARP QUANTIFICATION ANALYSIS: A FRAMEWORK FOR PATH-BASED SIGNAL ALIGNMENT METRICS

*Sir-Lord Wiafe*[1]*, Vince D. Calhoun*[1]

[1]Tri-Institutional Center for Translational Research in Neuroimaging and Data Science (TReNDS), Georgia State University, Georgia Institute of Technology, and Emory University, Atlanta, GA 30303, USA

## ABSTRACT

Dynamic time warping (DTW) is widely used to align time series evolving on mismatched timescales, yet most applications reduce alignment to a scalar distance. We introduce warp quantification analysis (WQA), a framework that derives interpretable geometric and structural descriptors from DTW paths. Controlled simulations showed that each metric selectively tracked its intended driver with minimal crosstalk. Applied to large-scale fMRI, WQA revealed distinct network signatures and complementary associations with schizophrenia negative symptom severity, capturing clinically meaningful variability beyond DTW distance. WQA transforms DTW from a single-score method into a family of alignment descriptors, offering a principled and generalizable extension for richer characterization of temporal coupling across domains where nonlinear normalization is essential.

## 1. INTRODUCTION

Many complex systems interact across mismatched timescales, such as variable speaking rates, irregular biomedical delays, and locally jittered sensors [1, 2]. Dynamic time warping (DTW) aligns such sequences via nonlinear time normalization [3]. Prior works have extensively focused on the DTW distance and key variants, including constrained windows [4], derivative DTW [5], weighted DTW [6], multiscale DTW [7], and soft-DTW [8]. While useful, the distance collapses alignment to a single scalar and obscures how normalization evolves.

Prior work has demonstrated that the warping path itself can yield richer insights. Studies have used it to identify lead–lag relationships between sequences [9], and more recently, the concept of warp elasticity quantified local rates of stretching and compression, revealing reproducible timing asymmetries in fMRI data[10, 11]. Yet there remains no systematic framework that treats the path as a structured object from which multiple, complementary descriptors can be derived.

Inspired by recurrence quantification analysis in dynamical systems [12], we propose warp quantification analysis (WQA): a family of interpretable metrics that decompose DTW paths into geometric descriptors (continuous deviations from the diagonal) and structural descriptors (discrete organizational features). This provides a structured characterization of alignment dynamics, offering information unavailable from DTW distance alone and applicable across diverse signal domains.

We validate WQA through controlled simulation scenarios that isolate each metric's unique sensitivity and further demonstrate its practical utility on large-scale fMRI data, where the metrics provide interpretable markers of cognitive variability beyond DTW distance.

## 2. METHODS

### 2.1. Dynamic time warping

Dynamic time warping is a standard technique for aligning two time series that may differ in rate or timing [3]. DTW finds an optimal monotone mapping between indices of two discrete time series $x = \{x_i\}_{i=1}^{N}$ and $y = \{y_i\}_{j=1}^{M}$ within a Sakoe–Chiba window of radius $w$. The alignment path is defined as follows:

$$\varphi(\tau) = \{(\varphi_x(\tau), \varphi_y(\tau))\}_{\tau=1}^{L}, \qquad \varphi \in \Pi, \tag{1}$$

where $\Pi$ denotes all admissible paths within the window $|i-j| \leq w$, and $L$ is the path length. The cumulative DTW distance cost is

$$D = \min_{\varphi\in\Pi} \sum_{\tau=1}^{L} \lambda_\gamma(x_{\varphi_x(\tau)}, y_{\varphi_y(\tau)}) \tag{2}$$

with generalized pointwise distance [13]

$$\lambda_\gamma(x_i, y_j) = |x_i - y_j|^\gamma, \quad \gamma > 0. \tag{3}$$

DTW thereby yields both a scalar distance $D$ and the optimal warping path $\varphi$.

### 2.2. Warp quantification analysis

We introduce WQA, a framework that derives interpretable descriptors from the DTW warping path rather than relying only on the cumulative distance. WQA formalizes the warping information into two complementary families of metrics: **geometric descriptors**, which quantify continuous deviations of the path from the diagonal (e.g., typical offset, variability, path-length overhead), and **structural descriptors**, which capture discrete features of path organization (e.g., persistence of diagonal segments, frequency of diagonal side-switching).

#### *2.2.1. Warp distortion ratio (WDR) — geometric*

The warp distortion ratio (WDR) quantifies how much longer the warping path is compared to the minimal diagonal. For a warping path of length $L$ aligning sequences of lengths $N$ and $M$, the minimal feasible path length is $\max(N, M)$ and the maximal is $N + M$. We define

$$WDR = \frac{L - \max(N, M)}{N + M - \max(N, M)} \tag{4}$$

As a geometric descriptor**,** WDR quantifies the relative path overhead: how much the alignment requires insertions or repeats beyond 1:1 matching. Higher values indicate stronger local distortions, while lower values reflect paths close to diagonal alignment.

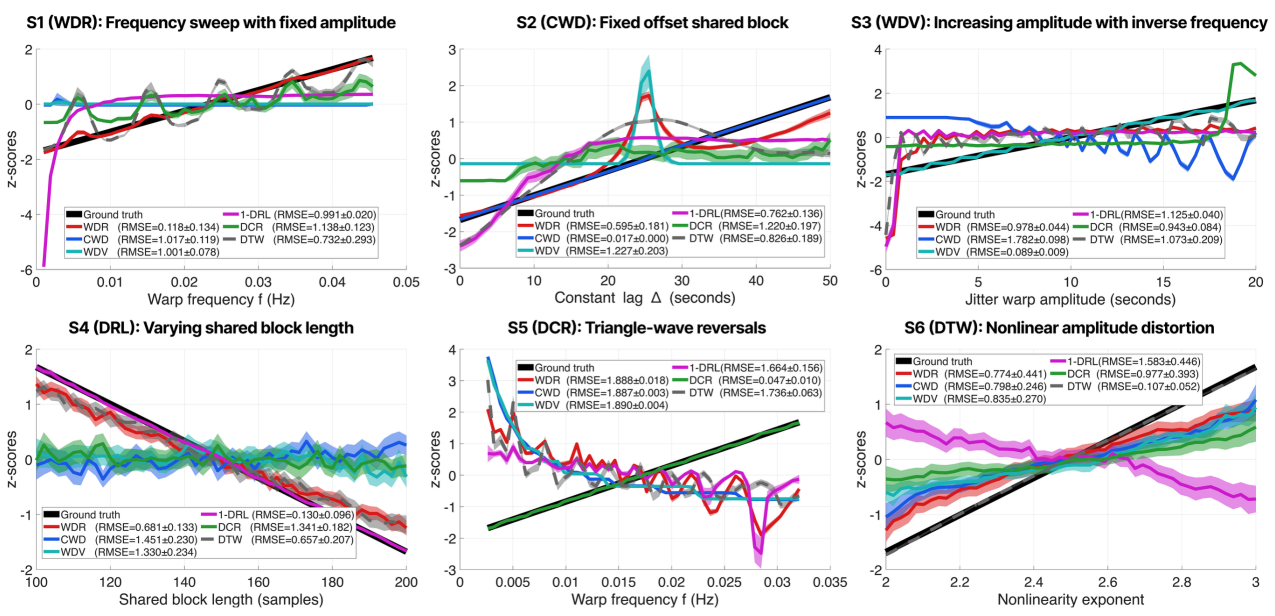

*Figure 1. Six controlled scenarios (S1–S6) isolate each metric's driver; curves show z-scored responses with 95% CIs across 1000 simulations, confirming selective sensitivity across metrics.*

### *2.2.2. Central warp deviation (CWD) — geometric*

The central warp deviation (CWD) measures the typical absolute offset of the warping path from the diagonal, reflecting a systematic temporal offset between the sequences. Formally, let

$$WD(\tau) = \varphi_x(\tau) - \varphi_y(\tau), \quad \tau = 1, \dots, L, \tag{5}$$

denote the warp deviation (WD) of the warping path from the diagonal at step $\tau$ [10]. Then CWD is defined as the central tendency of $|WD(\tau)|$, with the option to compute either the median or the mean:

$$CWD = \frac{1}{w} \cdot median\{|WD(\tau)|\}_{\tau=1}^{L} \tag{6}$$

Here, the window $w$ is used for normalization so that CWD lies in [0,1]. A mean version can be adopted for very long sequences.

As a geometric descriptor**,** CWD captures the typical temporal shift required for alignment. Low CWD means the path stays near the diagonal, with signals evolving nearly in sync. High CWD means the path is consistently displaced.

### *2.2.3. Warp deviation variability (WDV) — geometric*

The warp deviation variability (WDV) measures the spread or dispersion of the warping path's offset from the diagonal, i.e., how irregular or jittery the temporal alignment is once the two sequences have been matched. While CWD captures the *typical* offset, WDV captures how much that offset fluctuates along the path, as below:

$$WDV = \frac{1}{w} median\Big\{\big||WD(\tau)| - median(|WD(\tau)|)\big|\Big\}_{\tau=1}^{L} \tag{7}$$

Here, the window $w$ is used for normalization so that WDV lies in [0,1]. The standard deviation can be adopted for very long paths.

As a geometric descriptor**,** WDV measures the stability of alignment offsets. Low WDV indicates nearly constant displacement, while high WDV reflects irregular oscillations, jitter, or switching in local adjustments.

### *2.2.4. Diagonal run length (DRL) — structural*

The diagonal run length (DRL) measures the persistence of contiguous diagonal steps in the warping path, i.e., stretches where both indices advance together $(\Delta\varphi_x, \Delta\varphi_y) = (1, 1)$. Such runs indicate intervals of stable synchrony, where the two sequences evolve in lockstep without insertions or repeats. Formally, define the step indicators

$$d(\tau) = \begin{cases} 1 \text{ if } \ \varphi_x(\tau) - \varphi_x(\tau-1) = 1 \\ \text{and } \varphi_y(\tau) - \varphi_y(\tau-1) = 1, \\ \quad 0 \text{ otherwise} \end{cases} \qquad \tau = 2, \dots, L \tag{8}$$

Let the contiguous sequences of indices where $d(\tau) = 1$ have lengths $\{\ell_r\}_{r=1}^{R}$, where $R$ is the total number of such diagonal runs. After enforcing a dwell threshold $k$ to ignore short, spurious diagonals, we keep only those with $\ell_r \geq k$. The DRL is then defined as the central tendency of the retained run lengths, normalized by path length:

$$DRL = \frac{1}{L-1} median\{\ell_r : \ell_r \geq k\} \tag{9}$$

Here, $k \geq 1$ is the dwell parameter, and $L - 1$ is the maximum possible diagonal run length. Also, a mean version can be adopted for long sequences.

As a structural descriptor**,** DRL measures sustained synchrony. High DRL means the path contains long uninterrupted diagonals (coherent coupling), while low DRL means only short fragments (unstable or intermittent synchrony). For consistency with the other metrics (where larger values indicate greater disruption), in our further analyses, we use $1 - DRL$, so that higher values correspond to weaker synchrony.

### *2.2.5. Diagonal crossing rate (DCR) — structural*

The diagonal crossing rate (DCR) measures how frequently the warping path switches sides of the diagonal, i.e., sustained reversals in which sequence is ahead. Such crossings reflect alternations in temporal offset, where the alignment repeatedly flips between one signal running ahead and the other catching up. Formally, define the offset-sign sequence

$$s(\tau) = sign(WD(\tau)), \quad \tau = 1, \dots, L, \tag{10}$$

with $s(\tau) \in \{-1,0,1\}$. To avoid counting spurious $0 \to \pm1 \to 0$ micro-events as crossings, we form a zero-resolved sign sequence $\tilde{s}(\tau)$ by propagating the most recent nonzero sign forward:

$$\tilde{s}(\tau) = \begin{cases} s(\tau), & s(\tau) \neq 0, \\ \tilde{s}(\tau-1), & s(\tau) = 0, \end{cases} \tag{11}$$

Each contiguous sign forms a single run $r$ with a run-length encoding of $\{(v_r, \ell_r)\}_{r=1}^{R}$, where $v_r \in \{-1,1\}$ is the constant sign on run $r$, $\ell_r$ is its length, and $R$ the total number of runs. After enforcing a dwell threshold $k$, we count only those sign-flip boundaries where both adjacent runs persist for at least $k$ steps. A dwell-qualified crossing is a boundary $r \in \{1, \dots, R-1\}$ such that $v_r v_{r+1} = -1$ and $\ell_r \geq k, \ell_{r+1} \geq k$. The diagonal cross count is

$$C_k = \sum_{r=1}^{R-1} \mathbf{1}\,[v_r v_{r+1} = -1, \ell_r \geq k, \ell_{r+1} \geq k\,] \tag{12}$$

The normalized diagonal crossing ratio is then defined as

$$DCR = \frac{2k}{L-1} C_k \tag{13}$$

Here, $\frac{2k}{L-1}$ is the maximum possible crossing density given the dwell constraint. By construction, $C_k \leq \frac{L-1}{2k}$ so $DCR \in [0,1]$.

As a structural descriptor**,** DCR measures how often the warp deviation changes sign in a sustained way. Low DCR reflects stability on one side of the diagonal, while high DCR indicates repeated, persistent alternations. DCR complements DRL: long diagonal runs tend to reduce DCR, but the two are not strict inverses.

### 2.3. Simulation design and setup

To illustrate the distinct sensitivities of our WQA metrics, we designed controlled simulation scenarios in which synthetic signals were related through explicit warp functions. All simulations used synthetic signals of length $N = 1000$ with sampling time $T_s = 1s$, band-limited to $0.01 - 0.1\,Hz$. A base signal $x(t)$ was generated as a filtered Gaussian process, and a warped counterpart was produced by resampling the base signal at displaced times using shape-preserving interpolation:

$$y(t) = x(t'), \quad t' = t + u(t), \tag{14}$$

where $u(t)$ is a prescribed warp displacement function and $t'$ is the time warped function. DTW estimated the warping path $\varphi$ under a Sakoe–Chiba window of radius $w = 0.2N$. For each metric, we swept a single parameter across a safe range over 1000 simulations and estimated WDR, CWD, WDV, DRL, DCR, and DTW distance. Both ground-truth drivers and metric outputs were z-scored before comparison, and performance was quantified by root mean squared error (RMSE) of the mean curve with 95% confidence intervals.

*Scenario 1 (WDR):* To isolate path-length overhead, we used a lifted cosine displacement, $u(t) = \frac{A}{2}(1 - \cos(2\pi f t))$, which ranges in $[0, A]$. Amplitude $A$ is fixed while frequency $f$ is swept. Because the derivative of the time warped function is $\frac{dt'}{dt} = 1 + (A\pi f)\sin(2\pi f t)$, monotonicity is enforced by requiring $A\pi f < 1$. Increasing $f$ makes the warping path accumulate more extra steps (insertions/repeats), so WDR grows, while other metrics remain largely unrelated.

*Scenario 2 (CWD):* To isolate systematic displacement, we embed a shifted block within the signals. We select a block of length $P$ samples starting at index $s$, and insert it into $y(t)$ with a fixed offset, $\mu$: $y(t) = \begin{cases} x(t-\mu), & t \in [s, s+P], \\ \eta(t), & otherwise, \end{cases}$ where $\eta(t)$ is independent noise filtered in the same frequency band $[0.01 - 0.1\,Hz]$. The offset $\mu$ represents a fixed temporal shift between shared blocks. This ensures that the two signals share an identical contiguous segment of length $P$, but consistently displaced by $\mu$. Outside the block, there is no meaningful alignment. By sweeping $\mu$ while holding $P$ fixed, the warping path is forced to sit offset from the diagonal by a nearly constant amount, driving CWD while leaving other metrics comparatively unaffected.

*Scenario 3 (WDV):* To isolate alignment variability, we impose an oscillatory displacement with a fixed offset $\mu$: $u(t) = \mu + A\cos(2\pi f(A)t)$, $f(A) = c/A$ , where $A$ is the oscillation amplitude and $f(A)$ decreases inversely so that larger $A$ yields slower oscillations. $c$ is a fixed scale factor ensuring the oscillation frequency stays within the signal and window bandwidth. Sweeping $A$ increases jitter amplitude while $f(A) = c/A$ ensures oscillation slows proportionally, so the warping path oscillates more widely but does not violate monotonicity, thereby isolating WDV.

*Scenario 4 (DRL):* To isolate sustained synchrony, we construct signals that share an identical block of length $P$ starting from $s$, $y(s{:}s+P) = x(s{:}s+P)$ while keeping the rest as bandlimited unrelated Gaussian samples. Varying $P$ directly controls the persistence of uninterrupted diagonal runs, isolating DRL.

*Scenario 5 (DCR):* To isolate reversal frequency from confounding changes in offset or jitter amplitude, we use a triangle-wave displacement $u(t) = A\,\mathrm{tri}(2\pi f)$, where $\mathrm{tri}(\cdot)$ is the standard triangular wave. Its derivative is piecewise constant, $\frac{du}{dt} = \pm 4Af$, yielding strictly linear segments of forward and backward slope. Unlike a sinusoid, this ensures equal-time intervals on each side of the diagonal (50% duty cycle), producing symmetric, sustained reversals whose frequency is exactly controlled by $f$. Sweeping $f$ increases the crossing rate in a clean, interpretable manner, isolating DCR.

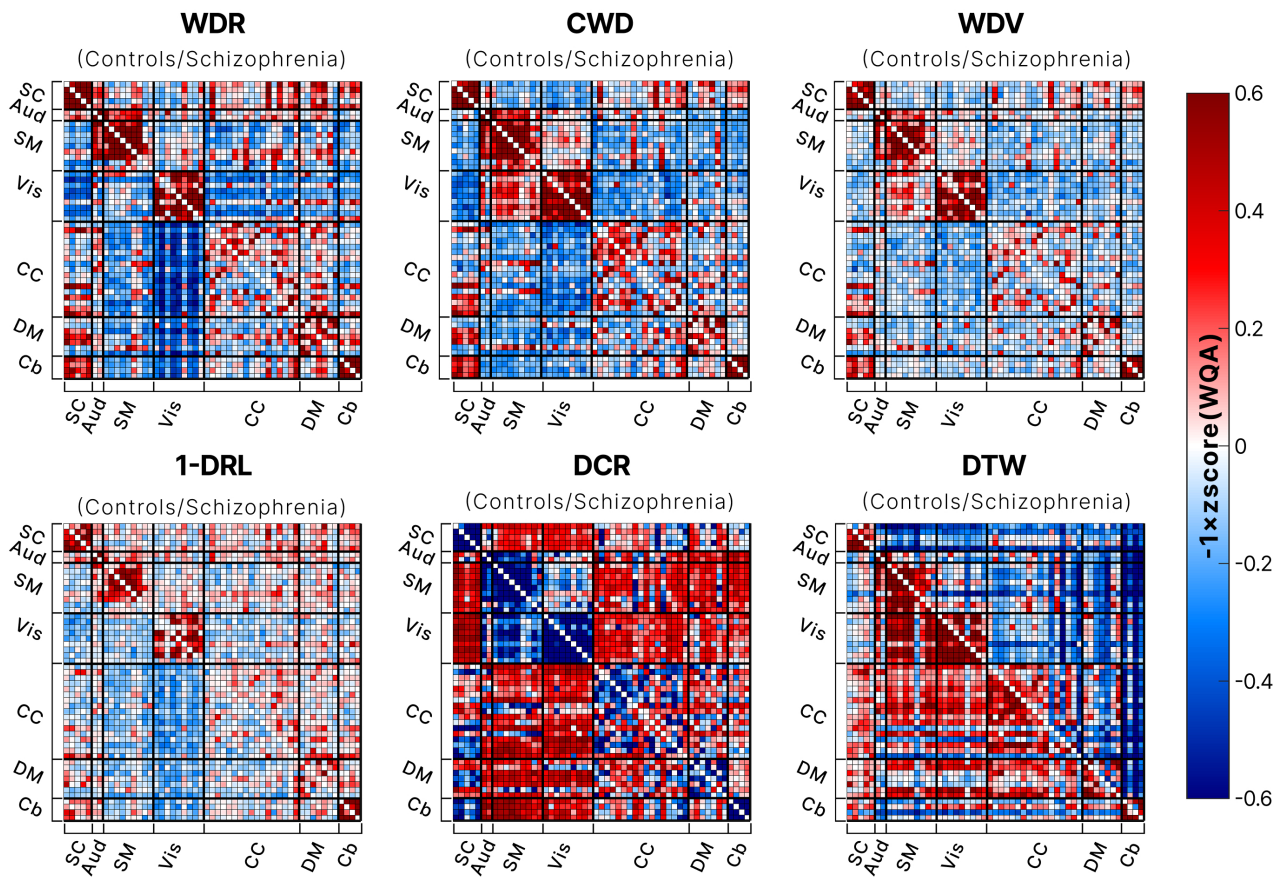

*Figure 2. Group-mean connectivity matrices for each WQA metric and DTW distance. Lower triangles: controls; upper triangles: schizophrenia. Values were z-scored and sign-inverted (–1×z-score) to mimic the conventional fMRI connectivity display, where positive values (red) indicate stronger coupling and negative values (blue) indicate weaker coupling. Each metric yields distinct connectivity patterns, highlighting complementary aspects of alignment dynamics that are not captured by the DTW distance.*

*Scenario 6 (DTW):* To show that the DTW distance also carries unique information, we isolated distributional mismatch by applying a nonlinear pointwise distortion without altering timing $y(t) = sign\big(x(t)\big) \cdot |x(t)|^{1+\alpha}$. By sweeping the exponent $\alpha$ only, the diagonal path is expected to be preserved as $u(t) = 0$ so all WQA metrics remain flat, while DTW distance rises systematically with distributional mismatch.

In our simulations, we set the DRL and DCR dwell threshold to $k = 3$. The DTW parameter was fixed at $\gamma = 1$ a value that balances sensitivity across amplitude scales while keeping the focus on time-warping functions rather than signal magnitude [13]. For CWD and WDV, we used the median-based definitions to enhance robustness to local fluctuations. These complementary stress tests demonstrate that each metric, including DTW distance, responds selectively to its intended driver while others remain unrelated, ensuring interpretability and non-redundancy. The full implementation details and reproducibility scripts are available in our open-source code repository (https://github.com/Sirlord-Sen/warp-quantification-analysis).

## 2.4. Dataset and data processing

We analyzed resting-state fMRI from the Function Biomedical Informatics Research Network (fBIRN) consortium. Data were acquired with a repetition time of 2 s and comprised 160 healthy controls and 151 individuals with. Data underwent standard preprocessing [14], and the NeuroMark independent component analysis (ICA) pipeline [15] was used to produce 53 intrinsic connectivity networks (ICNs). These ICNs were despiked, detrended, band-pass filtered (0.01–0.15 Hz), and z-scored before pairwise DTW and WQA analyses. This study was determined by the GSU IRB to be 'not human subjects' and thus does not require ethical approval.

Prior work in fMRI has established a 100-s window as a robust choice for DTW[16]; we therefore adopt the same setting for our analyses, corresponding to 50 samples at TR = 2 s. The DTW parameter in equation (3) was set to $\gamma = 2$, consistent with evidence of high test–retest reliability in fMRI and its direct mapping to the standard definition of signal energy disparity between DTW-aligned signals [17]. We set the DRL and DCR dwell thresholds to $k = 3$, reducing the risk of spurious outcomes.

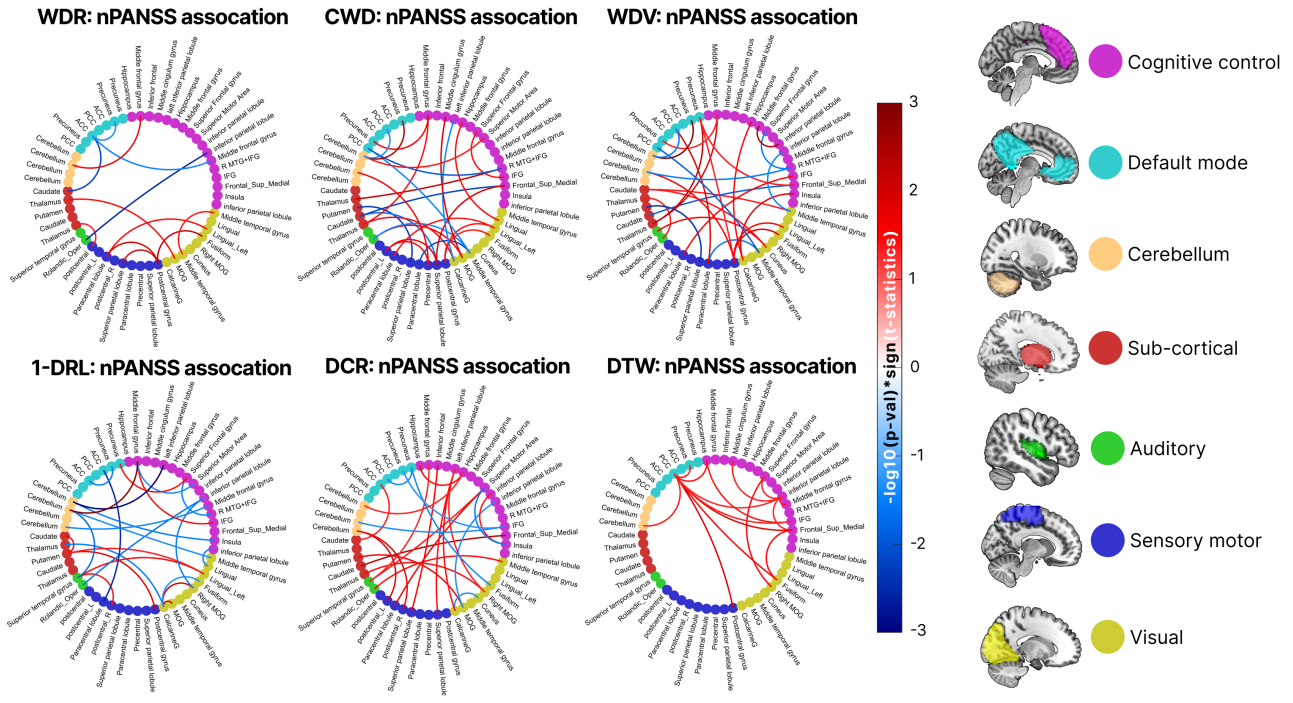

*Figure 3. Connectogram diagrams of FDR-corrected associations between connectivity pairs and negative symptom severity (nPANSS) in schizophrenia. Red connections indicate positive associations, and blue connections indicate negative associations. Each metric highlights a distinct set of significant region pairs, reflecting complementary sensitivity to different alignment dynamics.*

### 2.5. Real-world fMRI analysis

To show that the metrics capture distinct and clinically relevant patterns in real data, we performed group-level analyses and symptom association testing. Specifically, we computed group-averaged WQA and DTW matrices across subjects to visualize network-level patterns, then used a generalized linear model to test associations between negative symptom scores and individual metrics at each network pair, controlling for age, sex, mean frame displacement, and site. Results were corrected for multiple comparisons across metrics and network pairs using the false discovery rate (FDR).

## 3. RESULTS

### 3.1. Simulation results and metric sensitivities

Controlled simulations confirmed that each WQA metric selectively tracked its intended driver while performing poorly in other scenarios (Figure 1). In the path-length manipulation (S1), only WDR rose systematically with increasing warp frequency ($RMSE = 0.118 \pm 0.134$), vs. others ($RMSE \geq 0.732$). When a fixed offset was imposed (S2), CWD increased steadily with offset size ($RMSE = 0.017 \pm 0.001$), vs. others ($RMSE \geq 0.595$). Oscillatory displacements with growing amplitude but slowing frequency (S3) produced larger jitter, captured uniquely by WDV ($RMSE = 0.089 \pm 0.009$), vs. others ($RMSE \geq 0.943$). Extending shared blocks (S4) increased uninterrupted diagonals, with 1–DRL decreasing accordingly ($RMSE = 0.130 \pm 0.096$), vs. others ($RMSE \geq 0.657$). Triangle-wave reversals (S5) yielded proportional increases in DCR ($RMSE = 0.047 \pm 0.01$), vs. others ($RMSE \geq 1.664$). Finally, nonlinear pointwise distortions (S6) raised the DTW distance ($RMSE = 0.107 \pm 0.052$) against all WQA metrics ($RMSE \geq 0.774$), since temporal warping was absent. These outcomes show each metric selectively tracks its designed manipulation with minimal crosstalk, establishing interpretability and complementarity.

### 3.2. Real-world fMRI analysis results

Applying WQA and DTW to resting-state fMRI revealed distinct group-average connectivity patterns across brain networks (Figure 2). To aid interpretation, WQA values were z-scored across all subjects and network pairs and sign-inverted so that negative values correspond to higher WQA values and positive values to lower WQA values, matching the polarity convention of conventional fMRI connectivity displays. Although almost all metrics emphasized strong within-domain coupling, they diverged in emphasis. For example, WDR highlighted stronger path-length distortions between higher-order networks—cognitive control (CC), default mode (DM)—and visual regions (Vis), whereas DTW distance was dominated by a systematic group-level contrast between controls and schizophrenia. Notably, DCR revealed the opposite of the expected trend: within-domain networks, while strongly aligned, exhibited higher switching rates than cross-domain pairs. This suggests that even tightly coupled systems engage in rapid exchanges of dominance, a dynamic invisible to DTW distance and consistent with theories of competitive–cooperative interactions within functional domains, where subregions transiently alternate in driving network activity [18, 19].

We next tested associations with negative symptom severity in schizophrenia (Figure 3). FDR-corrected analyses revealed that different metrics isolated partially non-overlapping sets of significant region pairs, with both positive (red) and negative (blue) associations. Structural descriptors (1-DRL, DCR) highlighted widespread links including cross-domain interactions, whereas geometric descriptors (WDR, CWD, WDV) emphasized both within- and cross-domain effects. DTW distance showed a comparatively more restricted pattern, concentrated in cognitive control and default mode networks. These findings demonstrate that WQA captures clinically meaningful variability beyond DTW distance, reinforcing its interpretability and complementarity in real-world neuroimaging data.

## 4. CONCLUSION

We proposed warp quantification analysis (WQA), a structured framework for extracting complementary metrics from DTW paths. Unlike the traditional DTW distance, WQA decomposes alignment dynamics into geometric and structural descriptors, yielding interpretable markers of temporal coupling. Simulation experiments confirmed that each metric is uniquely sensitive to its designed manipulation, while real-world fMRI analyses demonstrated that WQA recovers distinct connectivity patterns and clinically meaningful associations not visible with DTW distance alone. By treating the warping path as a structured object, WQA transforms DTW from a single scalar tool into a family of alignment descriptors. Although our present validation focused on fMRI and controlled simulations, the framework is broadly applicable to speech, biomedical, and sensor signals, providing a general foundation for future work on the dynamics of temporal alignment.

## 5. ACKNOWLEDGEMENT

This work was supported by the National Institutes of Health (NIH) grant (R01MH123610) and the National Science Foundation (NSF) grant #2112455.

## 6. REFERENCES

1. Pittman-Polletta, B.R., et al., *Differential contributions of synaptic and intrinsic inhibitory currents to speech segmentation via flexible phase-locking in neural oscillators.* PLOS Computational Biology, 2021. **17**(4): p. e1008783.
2. Ten Oever, S. and A.E. Martin, *An oscillating computational model can track pseudo-rhythmic speech by using linguistic predictions.* Elife, 2021. **10**: p. e68066.
3. Sakoe, H. and S. Chiba, *Dynamic programming algorithm optimization for spoken word recognition.* IEEE transactions on acoustics, speech, and signal processing, 2003. **26**(1): p. 43-49.
4. Salvador, S. and P. Chan, *Toward accurate dynamic time warping in linear time and space.* Intelligent data analysis, 2007. **11**(5): p. 561-580.
5. Keogh, E.J. and M.J. Pazzani, *Derivative Dynamic Time Warping*, in *Proceedings of the 2001 SIAM International Conference on Data Mining (SDM)*. p. 1-11.
6. Jeong, Y.-S., M.K. Jeong, and O.A. Omitaomu, *Weighted dynamic time warping for time series classification.* Pattern Recognition, 2011. **44**(9): p. 2231-2240.
7. Müller, M., H. Mattes, and F. Kurth. *An efficient multiscale approach to audio synchronization*. in *ISMIR*. 2006.
8. Cuturi, M. and M. Blondel. *Soft-dtw: a differentiable loss function for time-series*. in *International conference on machine learning*. 2017. PMLR.
9. Stübinger, J. and D. Walter, *Using multi-dimensional dynamic time warping to identify time-varying lead-lag relationships.* Sensors, 2022. **22**(18): p. 6884.
10. Wiafe, S.-L., et al., *The dynamics of dynamic time warping in fMRI data: a method to capture inter-network stretching and shrinking via warp elasticity.* Imaging Neuroscience, 2024. **2**: p. 1-23.
11. Wiafe, S.L., et al. *Capturing Stretching and Shrinking of Inter-Network Temporal Coupling in FMRI Via WARP Elasticity*. in *2024 IEEE International Symposium on Biomedical Imaging (ISBI)*. 2024.
12. ECKMANN, J.-P., S.O. KAMPHORST, and D. RUELLE, *Recurrence Plots of Dynamical Systems*, in *Turbulence, Strange Attractors and Chaos*. p. 441-445.
13. Herrmann, M., C.W. Tan, and G.I. Webb, *Parameterizing the cost function of dynamic time warping with application to time series classification.* Data Mining and Knowledge Discovery, 2023. **37**(5): p. 2024-2045.
14. Penny, W.D., et al., *Statistical parametric mapping: the analysis of functional brain images*. 2011: Elsevier.
15. Du, Y., et al., *NeuroMark: An automated and adaptive ICA based pipeline to identify reproducible fMRI markers of brain disorders.* Neuroimage Clin, 2020. **28**: p. 102375.
16. Meszlényi, R.J., et al., *Resting state fMRI functional connectivity analysis using dynamic time warping.* Frontiers in neuroscience, 2017. **11**: p. 75.
17. Wiafe, S.-L., et al., *Aberrant Recovery of Timescale-Aligned Amplitude Balance Links to Symptoms and Cognition in Schizophrenia.* bioRxiv, 2025: p. 2025.03.20.644399.
18. Fornito, A., et al., *Competitive and cooperative dynamics of large-scale brain functional networks supporting recollection.* Proceedings of the National Academy of Sciences, 2012. **109**(31): p. 12788-12793.
19. Koshino, H., et al., *Cooperation and competition between the default mode network and frontal parietal network in the elderly.* Front Psychol, 2023. **14**: p. 1140399.